\documentclass[showpacs,reprint,preprintnumbers,amsmath,amssymb,superscriptaddress,longbibliography,nofootinbib]{revtex4-2}

\usepackage[utf8]{inputenc}
\usepackage{amsmath}
\usepackage{amssymb} 
\usepackage{enumerate}
\usepackage{multirow}
\usepackage{braket}
\usepackage[toc,page]{appendix}
\usepackage{url}
\usepackage{graphicx}
\usepackage[pdftex,bookmarks,linktocpage,pdfpagelabels,plainpages=false,hyperfigures,linkcolor=blue,citecolor=blue]{hyperref} %for hypertext links
\hypersetup{colorlinks=true}

\usepackage{mathrsfs,amssymb}  
\usepackage{cancel}
\usepackage[normalem]{ulem}
\usepackage{cleveref}
\usepackage{tikz}
\usetikzlibrary{decorations.markings}
\usetikzlibrary{positioning}
\usetikzlibrary{shapes}
\usetikzlibrary{calc}

\newcommand{\beq}{\begin{equation}}
\newcommand{\eeq}{\end{equation}}

\newcommand{\bea}{\begin{eqnarray}}
\newcommand{\eea}{\end{eqnarray}}

\newcommand{\GeV}{{\rm GeV}}
\newcommand{\TeV}{{\rm TeV}}

\begin{document}

\title{Riemannian Data preprocessing in Machine Learning \\ to focus on QCD color structure} 
\author{A. Hammad}
\email{ahhammad@cern.ch}
\affiliation{Institute of Convergence Fundamental Studies, Seoultech, Seoul 01811, Korea}
\author{Myeonghun Park}
\email{parc.seoultech@seoultech.ac.kr}
\affiliation{Institute of Convergence Fundamental Studies, Seoultech, Seoul 01811, Korea}
\affiliation{School of Natural Sciences, Seoultech, Seoul 01811, Korea}
\affiliation{School of Physics, KIAS, Seoul 02455, Korea}
\date{Sept.\,08, 2022}

\begin{abstract}
Identifying the quantum chromodynamics (QCD) color structure of processes provides additional information to enhance the reach for new physics searches at the Large Hadron Collider (LHC). 
Analyses of QCD color structure in the decay process of a boosted particle have been spotted 
as information becomes well localized in the limited phase space. 
While these kind of a boosted jet analyses provide an efficient way to identify a color structure, 
the constrained phase space reduces the number of available data, resulting in a low significance. 
In this letter, we provide a simple but a novel data preprocessing method using a Riemann sphere to utilize a full phase space by decorrelating QCD structure from a kinematics.
We can achieve a statistical stability by enlarging the size of testable data set with focusing on QCD structure effectively.  We demonstrate the power of our method at the 
finite statistics of the LHC Run 2. Our method is complementary to conventional boosted jet analyses in utilizing QCD information over the wide range of a phase space.
 \end{abstract}

\maketitle
\noindent{\bf Introduction.}  A machine learning (ML) paradigm has been dominating in various data science fields as an automated feature engineering from raw data becomes possible.
Recent developments of computer hardwares enable ML to learning patterns from ``big data" with  heuristic and empirical approaches in designing neural networks.
The price for these developments is the exponentially increasing number of required parameters to squeeze a performance from big-data, for example $\mathcal{O}(10^{11})$ number of parameters in GPT-3 \,\cite{brown2020language}.  
As ML gets a complicated structure,  it becomes opaque to understand how ML concludes an answer.  

In the field of high energy physics, we have accumulated precise understandings in the Standard Model of a particle physics (SM)
due to the big data created at the controlled environment of the LHC.
The big data from the LHC, mostly from QCD processes has two faces, an invaluable asset in understanding the nature of complicated QCD dynamics and an obstacle in the search of a new physics. 
To cope with challenges in both directions, various state-of-the-art  ML algorithms have been aggressively utilized\,\cite{Feickert:2021ajf, Radovic:2018dip, Karagiorgi:2022qnh}.
But unlike commercial usages of ML, scientific researches severely require both interpretability and explainability  to be away from unknown systematics due to the ``black box" nature of ML.
To achieve above criteria, one can try either (1) to choose physics-inspired features as inputs instead of relying on automated feature extraction from raw data\,\cite{Komiske:2017aww,Bradshaw:2022qev} or 
(2) to design a method to interpret how ML works\,\cite{Chang:2017kvc, Jung:2019iii, Faucett:2020vbu}.

In this letter, we introduce a data preprocessing method for ML to focus on the QCD phenomena more efficiently to reduce the complexity of a neural network. 
Utilizing characteristics of QCD provides complementary information to analyses on the kinematics of a phase space. 
For example, while $N$-prong substructure identifies the phase space of a jet (top quark tagging\,\footnote{We would like to refer \cite{Plehn:2011tg} for the review at the early state of the LHC, and \cite{Kasieczka:2019dbj} as a good summary of ML applications.}), 
the pattern of soft QCD radiations from an energetic particle provides a clue about $SU(3)_c$ representation of that particle (quark $q$ v.\,s.\,gluon $g$ jet tagging\,
\cite{Jones:1988ay,Lonnblad:1990bi,Ellis:1992qq,Gallicchio:2011xc, Gallicchio:2011xq, Gallicchio:2012ez, Larkoski:2014pca, FerreiradeLima:2016gcz, Frye:2017yrw, Davighi:2017hok, Metodiev:2018ftz, Komiske:2018vkc, Larkoski:2019nwj,Dreyer:2021hhr}).
More interestingly the pattern of soft gluon radiations of particles from a decay process are related to the way of contracting color indices, which has been known as a ``color-flow".
By utilizing information from the color-flow, one can identify the status of a decaying particle under $SU(3)_c$  (singlet v.\,s.\,non-singlet tagging\,\cite{Gallicchio:2010sw, Hook:2011cq, Soper:2011cr, Curtin:2012rm, Lim:2018toa, Lin:2018cin, Chakraborty:2019imr, Kim:2019wns, Buckley:2020kdp}). 
Most of analyses on the color-flow have been focused on a ``local" region where QCD activities are confined within a small area formed by a boosted object. 
Main reason for this focus is to distinguish a jet formed by the hadronic decay of a boosted electroweak particle ($h/W^\pm/Z$) from ordinary QCD jets ($q/g$).
But in the current situation without any clues on a new physics at the LHC,
we need to increase the size of data for analyses by extending the region of interests to a ``resolved region" where particles from a decay process can be reconstructed 
separately.  To disentangle the color-flow information from the kinematics of a decaying particle, we introduce an inverse stereographic projection of 
the image of energy deposits to a Riemann sphere. 
For a proof of concept, we demonstrate our method in distinguishing a color singlet particle $h$ from a  color octet $\sigma$ with the finite statistics of the LHC Run-2. 

%%%%%%%%%%%%%%%%%
\begin{figure}[t!]
\centering{\includegraphics[width=0.45\textwidth]{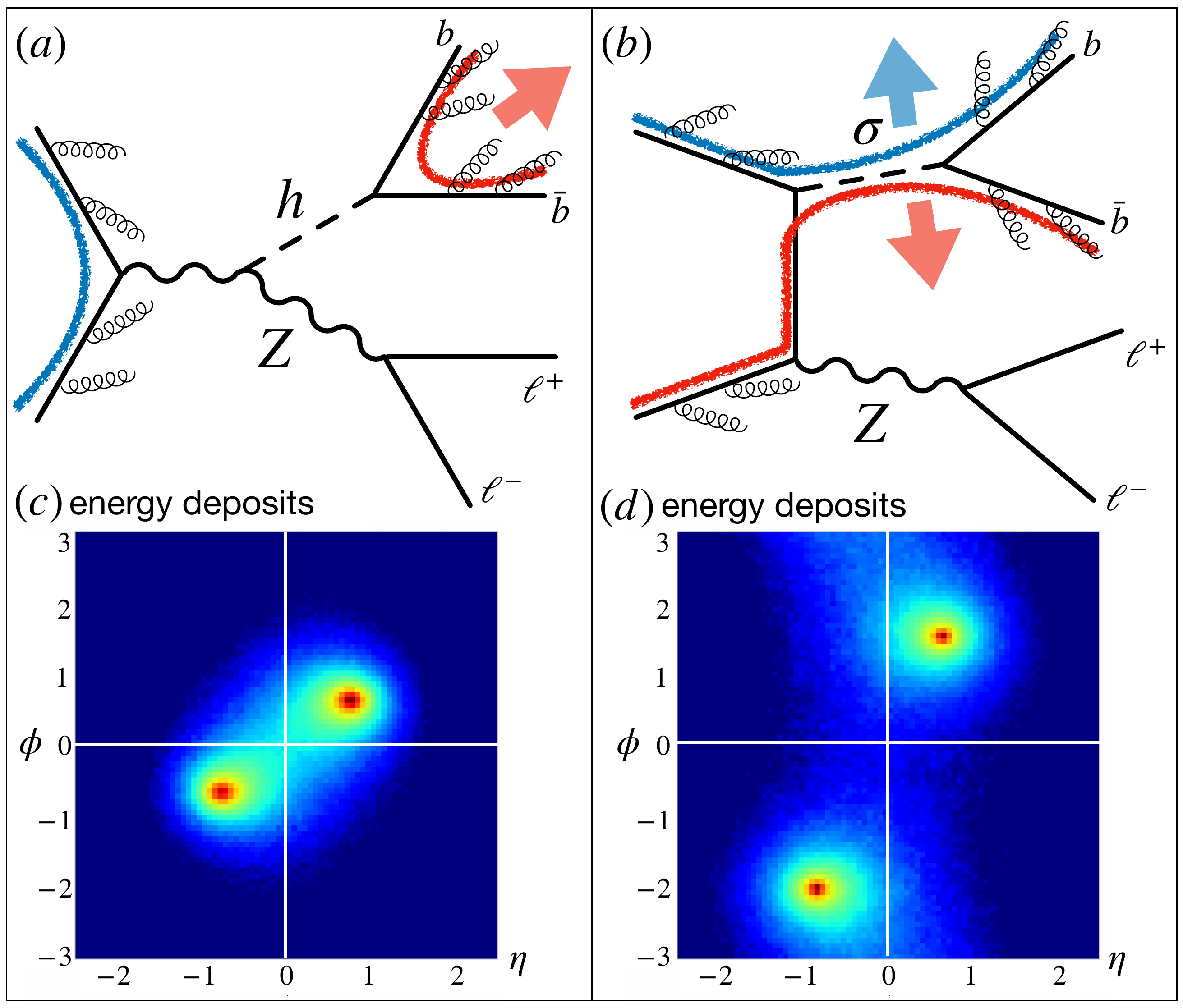}}
\caption{Feynman diagram for (a) color singlet $h$ and $Z$ production, and  (b) color octet $\sigma$ with $Z$ production. Here ``thick" color lines indicate the contractions of color indices of particles 
and big arrow for the tendency of soft gluon radiations. Density histogram of energy deposits from gluon radiations
 in (c) $h\to b\bar b$  and (d) $\sigma \to b \bar b$. Color code: high $p_T$ (red) to a low $p_T$ (blue) of particles.}
\label{fig:diagram}
\end{figure}
%%%%%%%%%%%%%%%%%
\noindent{\bf Riemannian preprocessing.}  
In Fig.\,\ref{fig:diagram}, we present the image of energy deposits either in the decay process of a color singlet particle $h$ to $b\, \bar b$ or a color octet particle $\sigma$ to $b\, \bar b$. Here $h$ and $\sigma$ have the same mass of $125\,\GeV$. To illustrate QCD radiation patterns with temperature plots, we apply multiple FSR/hadronization processes to a single partonic event. 
As we expect, soft gluon radiations from the decay process of a color singlet $h$ fill the internal space between $b$ and $\bar b$.
In the decay process of a color octet particle $\sigma$,  there are few energy deposits in the central region but towards a high rapidity region. 
This difference in QCD radiation patterns among different color states is clear when the angular distance $\Delta R_{(b\bar b)}$ between $b$ and $\bar b$ is fixed as Fig.\,\ref{fig:diagram}.
The angular distance $\Delta R_{(b\bar b)}$ is related to the energy of a decaying particle.  As the energy of a decaying particle increases, $\Delta R_{(b\bar b)}$ decreases.
%%%%%%%%%%%%%%%%%
\begin{figure}[h!]
\centering{\includegraphics[width=0.25\textwidth]{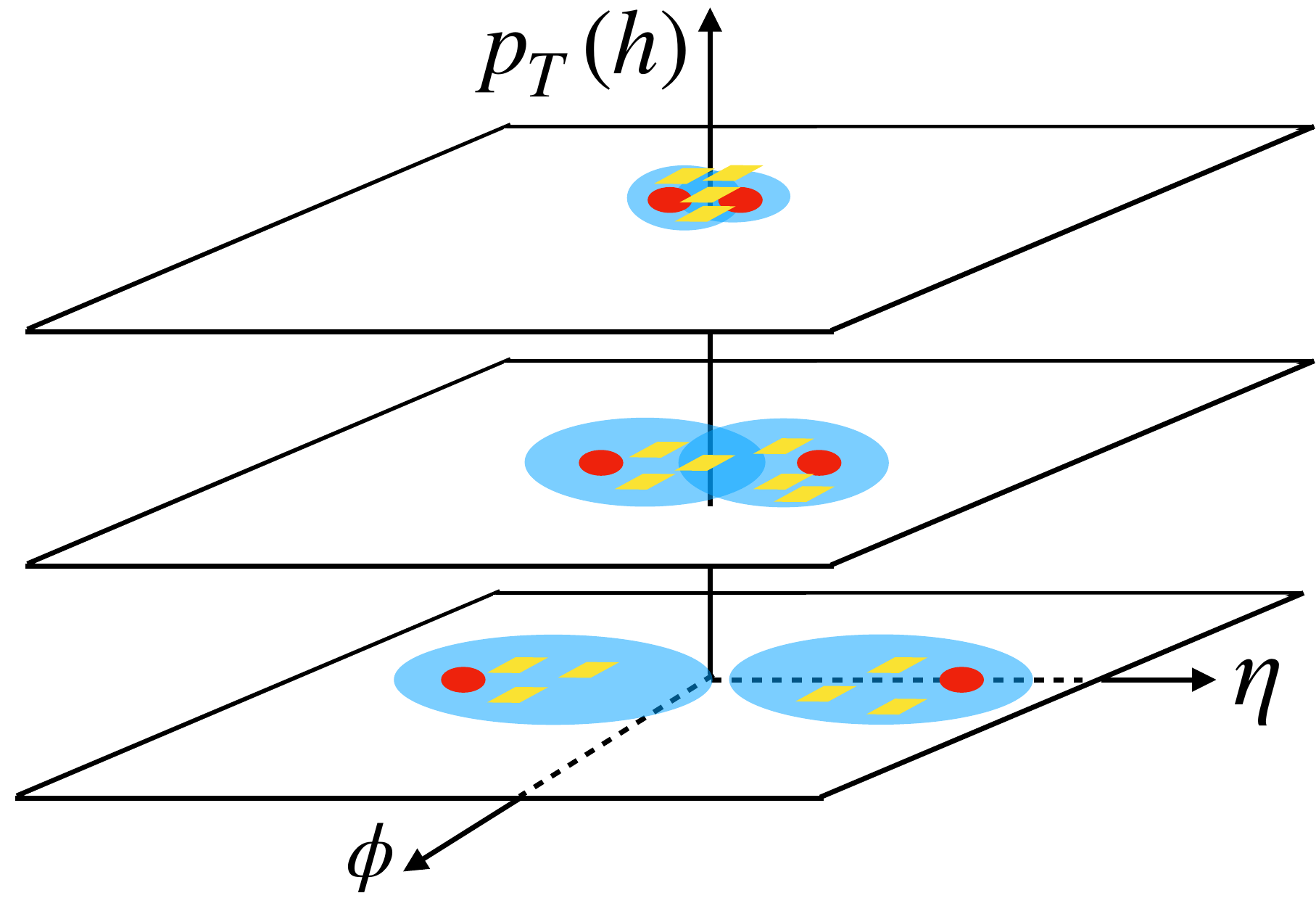}}
\caption{Energy deposits from the decay of $h$ according to $p_T$ of $h$. Here large energy deposits (red) are from $b$ and $\bar b$ quarks and soft energy deposits (yellow) are from gluon radiations. }
\label{fig:feature_space}
\end{figure}
%%%%%%%%%%%%%%%%%
%%%%%%%%%%%%%%%%%
\begin{figure}[t!]
\centering{\includegraphics[width=0.48\textwidth]{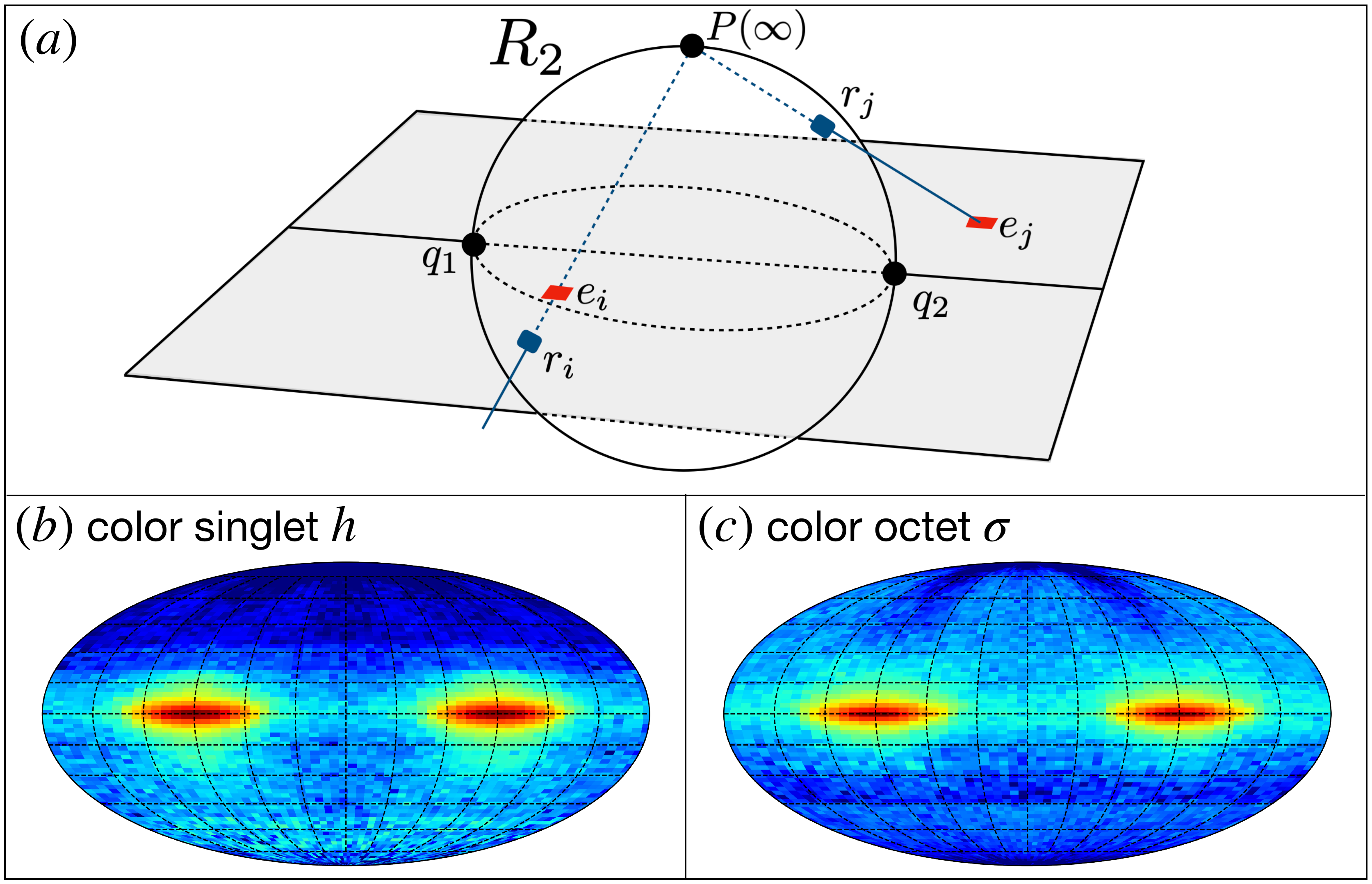}}
\caption{(a) The inverse stereographic projection of energy deposits. In $(\eta, \phi)$ plane, two largest energy deposits $q_1$ and $q_2$ determine a Riemann sphere $R_2$ 
with $\overline{q_1 q_2}$ as a diameter. The energy deposit $e_i$ inside $R_2$ is projected on the southern hemisphere at $r_i$ while the projection $r_j$ of an energy deposit $e_j$ outside $R_2$ locates on the northern hemisphere. We present the Mollweide maps of $R_2$ for energy deposits in the case of (b) $h\to b\bar b$ and (c) $\sigma \to b \bar b$. }
\label{fig:RProjection}
\end{figure}
%%%%%%%%%%%%%%%%%
In Fig.\,{\ref{fig:feature_space}, we illustrate how the image of energy deposits changes according to the $p_T$ of a decaying particle ($h$). 
Since energy deposits of $b$ and $\bar b$ are larger compared to the nearby soft gluon radiations, a priority will be given on varying distance between hot cores (red dots) rather than soft radiations in the ML training processes. Thus to keep the attention of ML training processes to the pattern of soft radiations, we introduce a data preprocessing procedure with an inverse stereographic projection as in Fig.\,{\ref{fig:RProjection} (a). Here $q_1$ and $q_2$ are the centers of each b-tagged jet. With shifting center of $(\eta, \phi)$ plane to the center $O_R$ of $\overline{q_1 q_2}$, 
we project an energy deposit $e_i$ to $r_i$ on the $R_2$ by extending a line from the north pole $P(\infty)$ of $R_2$ to $e_i$. The polar angle $\theta^R_i$ and azimuthal angle $\phi_i^R$ are determined as 
\beq
\cos\theta^R_i =  \left(\frac{\eta_i^2 + \phi_i^2 -r_R^2}{\eta_i^2 + \phi_i^2 +r_R^2}\right),  \quad
\phi^R_i = \arctan\left(\frac{\phi_i}{\eta_i}\right)\, .
\label{eq:Riemann}
\eeq
$r_R$ is the radius of a Riemann sphere $R_2$, which is related to an angular distance $\Delta R$ between $q_1$ and $q_2$ by $r_R=\Delta R/2$. Thus $r_R$ is related to the momentum of $b\bar b$ system. 
Here $\phi_i$ and corresponding $\phi_i^R$ are measured counterclockwise from $\overline{O_R q_2}$. 
To present projected images on $R_2$  to a two-dimensional space, we use a Mollweide projection, with locating $q_1$ and $q_2$ at the positions of  $-\frac{\pi}{2}$ and $\frac{\pi}{2}$ on an equator.  
As we observe in Fig.\,{\ref{fig:RProjection}(b) and (c), energy deposits from the decay of a color singlet particle $h$ are projected to the southern part of $R_2$ 
while energy deposits from the decay of a color octet $\sigma$ are mapped on the northern hemisphere.  
With considering only angular position $(\theta^R, \phi^R)$ on $R_2$, 
the patten of QCD soft radiations is disentangled from the energy of a decaying particle which is related to the radius of $R_2$. We recast the problem of identifying the color charge of a decaying particle into the problem of a simple binary pattern classification,  answering which part of  hemispheres is mostly populated. 

%%%%%%%%%%%%%%%%%
\begin{figure}[t!]
\centering{\includegraphics[width=0.48\textwidth]{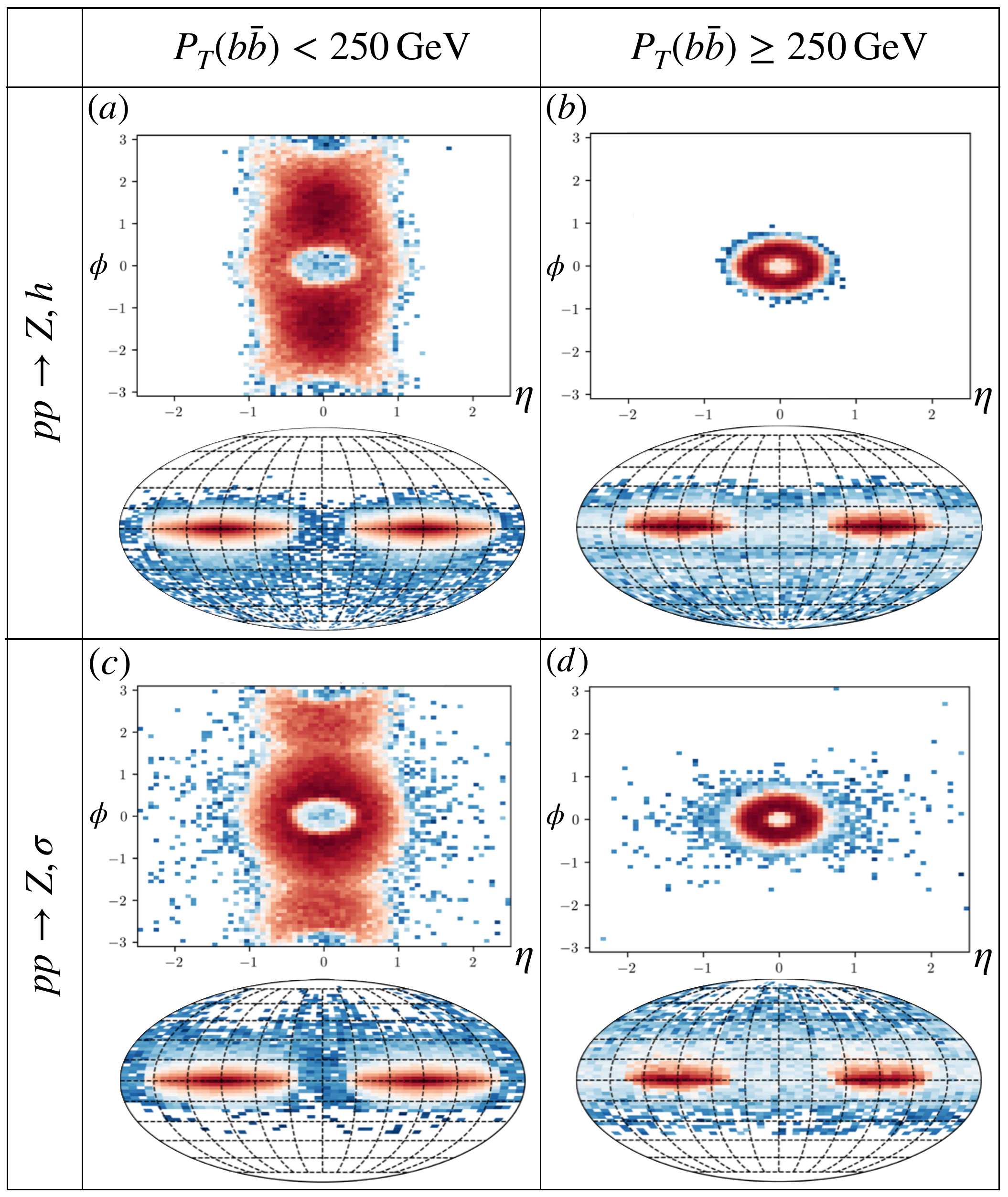}}
\caption{
We show energy deposits of particles except leptons from $Z$ at the detector level, dividing into four cases by particle case ($h$ or $\sigma$) and by $p_T(b\bar b)$.  In each slot, the upper rectangular plane is a calorimeter image with shifting center of two $b$-jets to the origin of $(\eta, \phi)$ plane while the lower earth-map type image is after Riemannian preprocessing.
For illustration purpose, we shots 15,000 events in each case with lowering the energy threshold of calorimeter to $0.8\GeV$.}
\label{fig:stackedPT}
\end{figure}
%%%%%%%%%%%%%%%%%

\noindent{\bf LHC example}  We test the performance of our proposal under the LHC environment with the standard chain of Monte Carlo simulations, {\sc MadGraph5}, {\sc Pythia8} and {\sc Delphes3} with {\sc Fastjet}\,\cite{Alwall:2014hca, Sjostrand:2014zea, deFavereau:2013fsa, Cacciari:2011ma}. We use MLM matching for an extra hard initial state radiation\,\cite{Mangano:2006rw}.
We consider $\sqrt{s}=13\TeV$ with an integrated luminosity $\mathcal{L}=139\,\rm{fb}^{-1}$ of the LHC Run 2. 
We follow event selection cuts and the corresponding number of signal events after cuts according to the ATLAS study on a Higgs and $Z$ boson with the channel of three or more jets and two opposite sign same flavor leptons\,\cite{ATLAS:2020fcp}. Event selection criteria are (1) three or more anti-$k_t$ jets of $R=0.4$ with $p_T >20\,\GeV$, (2) exactly two $b$-tagged jets with at least one $b-$jet's $p_T>45\,\GeV$, (3) exactly two leptons with $p_T>27\,\GeV$ with $81\,\GeV<m_{\ell\ell}<101\,\GeV$, and (4) the transverse momentum of a reconstructed $Z$-boson $p_T^V > 75\,\GeV$. 
The number of signal events we consider is 219, which is $1\sigma$ upper-deviation from the expected number of signals. For the alternative hypothesis of a signal, we take a color octet particle $\sigma$ as in FIG.\,\ref{fig:diagram} with the same selection cuts and number of events after cuts. 

To extract information from the image of energy deposits on the LHC sub-detectors, we utilize a basic Convolutional Neural Network (CNN) structure. 
We present an overview of training steps here, leaving detailed descriptions of our CNN structure in an appendix\,\ref{sec:appendix1}.
We prepare an image by adding energy deposits in sub-detectors of tracking, e-cal and h-cal within a window of $|\eta|<2.5$ and $|\phi|<\pi$ after correcting position of charged particle through the particle flow algorithm\,\cite{CMS:2017yfk}. 
The resulting image is discretized into $50\times 50$ pixels grid. 
Before we feed images into CNN model for a training, we take standard image preprocessing procedures including image cleansing, centering and normalization. 
To test the impact of our Riemannian preprocessing on CNN, we prepare two types of image data set, 
\begin{enumerate}
\item{\bf No preprocessing:} ordinary rectangular image in $(\eta, \phi)$ plane.
\item{\bf Riemannian preprocessing:} Mollweide image after Riemannian preprocessing as described in eq.\eqref{eq:Riemann} and  FIG.\,\ref{fig:RProjection}.
\end{enumerate}
We use the common CNN model to both types of image data set to check a performance change through ``Riemannian preprocessing" compared to ``No preprocessing" image data.

For a benchmark study, we also compare Riemannian preprocessed CNN with Lund jet plane analysis\,\cite{Khosa:2021cyk}. While our method is applied to the phase space of resolved jets, Lund jet plane analysis is focused on a boosted region where QCD activities from the decay process of a massive particle are confined within the narrow area to be captured by a single jet. 
Lund jet analysis is another type of preprocessing which transfers raw data of $(\eta, \phi)$ plane to a ``physically-motivated" plane of $(\ln\frac{1}{\triangle}, \ln\frac{k_t}{\GeV})$\,\cite{Dreyer:2018nbf}. 
Here $\triangle$ and $k_t$ are an angular distance and the transverse momentum of a soft emission with respect to a primary particle. 
For MC study, we follow selection cuts in \cite{Khosa:2021cyk} which requires (1) clustering jets with anti-$k_t$ with $R=1.0$, 
leading jet's (2) $p_T>250\,\GeV$, and (3) mass range of $110\,\GeV<m_J< 140\,\GeV$ on our MC data. The number of events after these selection cuts is 18. 
After passing cuts, we recluster a leading jet with C/A algorithm and use a Lund generator module of {\sc FASTJET contribute} code to construct a primary Lund jet plane. In this case, the dimension of an image has $25\times 25$ pixel size which is smaller than the image of a resolved region. Thus we adopt the CNN hyper-parameters in \cite{Khosa:2021cyk} which are tuned for a Lund jet image.
%%%%%%%%%%%%%%%%%
\begin{figure}[t!]
\centering{\includegraphics[width=0.40\textwidth]{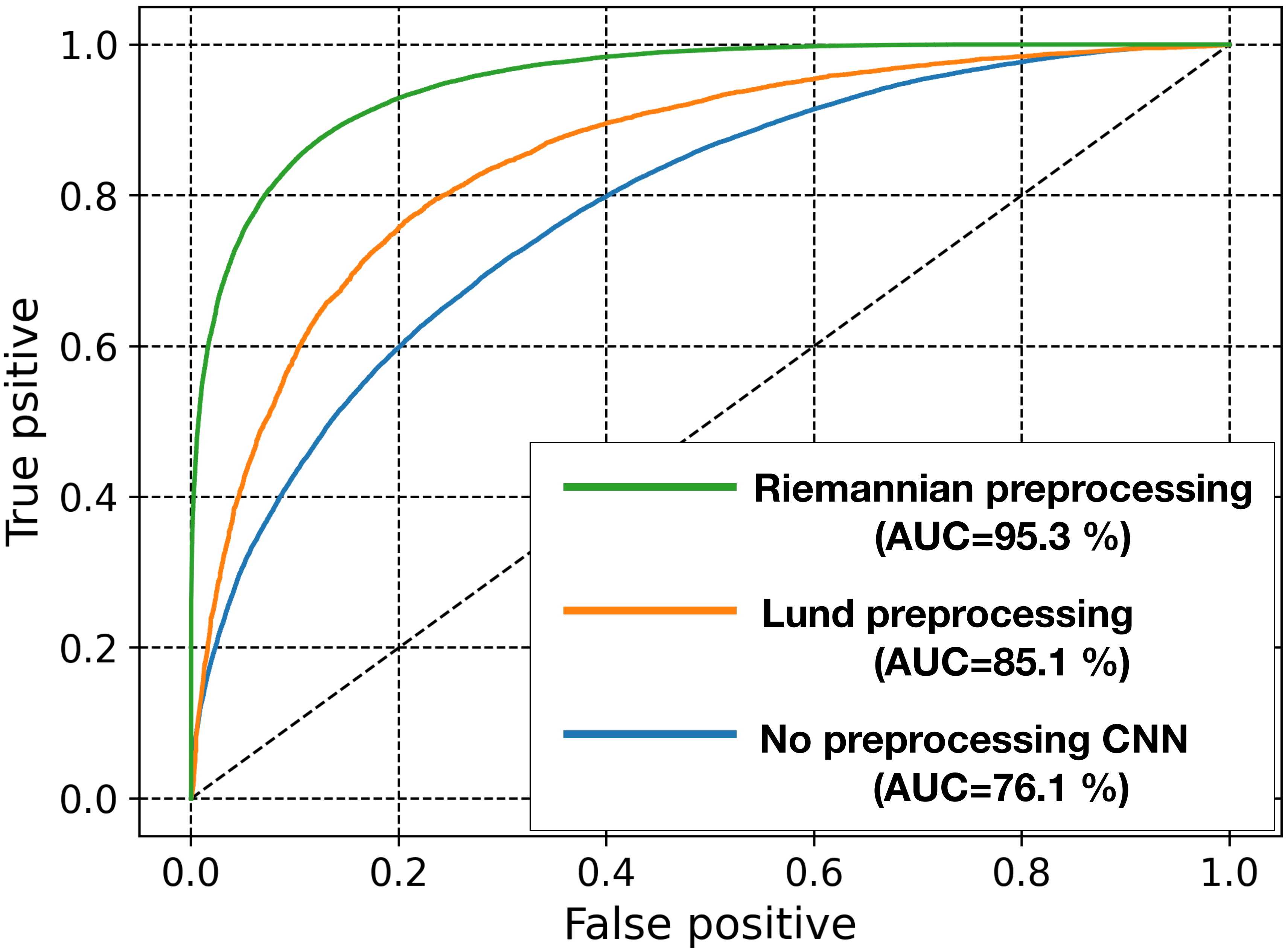}}
\caption{ROC curve from ML analyses with various data preprocessing method in the hypothesis test between a color singlet higgs $(h)$ and a color octet particle $\sigma$.}
\label{fig:ROC}
\end{figure}
%%%%%%%%%%%%%%%%%

To optimize CNN performance with different data preprocessing processes, 
we prepare two categories of $10^5$ events after selection cuts for (1) a resolved region with /without Riemannian preprocessing and for (2) boosted region for a Lund jet analysis. 
These MC samples are divided into $60\%, 20\%, 20\%$ for training, validation and test processes. 
We construct a receiver operating characteristic (ROC) curve to check  a separation power between different hypotheses (color singlet v.\,s.\,color octet)\,in FIG.\,\ref{fig:ROC}.
We observe that Riemannian preprocessed CNN analysis enhances the separation performance by $25\%$ compared to a conventional CNN analysis on raw images (No preprocessing CNN) in a resolved phase space region. 
As the phase space region where a Lund jet analysis is applied is totally disjoint from the one of the Riemannian preprocessing, we can not directly compare the performance between two analyses.  
What we need to pay attention is the statistics that each analysis can be applied. As we are limited by the number of data LHC creates, we should collect as many as events for a statistically stable analysis. 
We summarize the number of signal events for each analysis in Tab.\,\ref{tab:statistics}. As we see, we can utilize more data by the factor of $\mathcal{O}(10)$ in a resolved region compared to the boosted analysis. 
%%%%%%%%%%%%%%%%%
\begin{table}[h]
\begin{tabular}{|c|c|c|c|}
\hline
Phase space & cuts & Data Preprocessing method & $N_{\rm{sig}} $\\ \hline
Resolved region& Ref.\,\cite{ATLAS:2020fcp}   & Riemannian  & 219 \\ \hline
Boosted  region& Ref.\,\cite{Khosa:2021cyk}   & Lund jet plane & 18 \\ \hline
\end{tabular}
\caption{The number of signal events ($pp\to hZ\ \to b\bar b \ell \bar\ell$) after cuts in each analysis at the LHC with $\sqrt{s}=13\TeV$ and $\mathcal{L}=139\textrm{fb}^{-1}$. 
For each region, various kinematic cuts are optimized to suppress backgrounds.
 }
\label{tab:statistics}
\end{table}
%%%%%%%%%%%%%%%%%

To accurately compare the CNN classification performance, we carry out a test statistics using a log-likelihood ratio according to\,\cite{DeRujula:2010ys}.
Using a large number of test samples in a ML classification, we construct a probability distribution function (PDF) $P_{\mathbb{H}_{0,1}}\equiv P(\vec X| \mathbb{H}_{0,1})$ from a deep-learning score distribution of each hypothesis $\mathbb{H}_i$,  $\mathbb{H}_1$ for a color singlet particle and $\mathbb{H}_0$ for a color octet. Once we obtain PDFs, we define a log-likelihood ratio as 
\beq
\Lambda = \ln\frac{\mathcal{L}(\mathbb{H}_1)}{\mathcal{L}(\mathbb{H}_0)} =  \ln\frac{\Pi_i\,P_{\mathbb{H}_1}(\vec X_i)}{\Pi_i\,P_{\mathbb{H}_0}(\vec X_i)}
=\sum_{i=1}^{N_\textrm{evt}}\ln\ \frac{P_{\mathbb{H}_1}(\vec X_i)}{P_{\mathbb{H}_0}(\vec X_i)}, 
\eeq
here $N_{\textrm{evt}}$ is a number of events\footnote{There are studies on a statistical analysis using ML\,\cite{Coccaro:2019lgs, Khosa:2022vxb, Arganda:2022qzy}.}.
%%%%%%%%%%%%%%%%%
\begin{figure}[t!]
\centering{\includegraphics[width=0.5\textwidth]{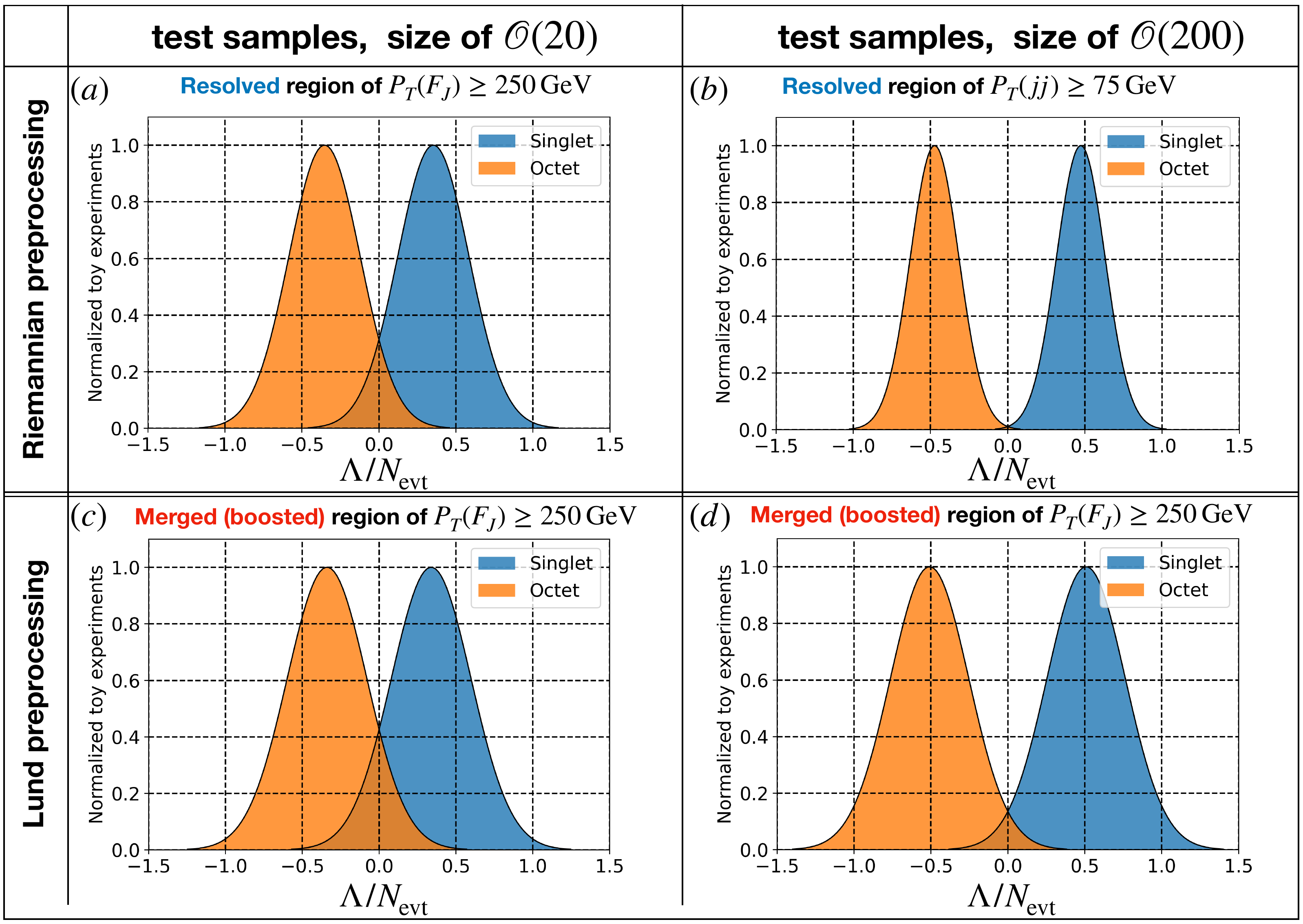}}
\caption{Log-likelihood ratio tests for the case of $(a), (b)$ Riemann preprocessing  and $(c), (d)$ Lund preprocessing.}
\label{fig:pseudoexp}
\end{figure}
%%%%%%%%%%%%%%%%%
To construct PDFs $P(\Lambda | \mathbb{H}_{0,1})$, we create 1M pseudo-experiments each for the case of a boosted region and a resolved region.
FIG.\,\ref{fig:pseudoexp} shows distributions of a log-likelihood ratio $\Lambda / N_{\textrm{evt}}$ with fixed $N_{\textrm{evt}}$ over pseudo-experiments. 
To quantify a separation power between different hypotheses, we calculate Type I error $\alpha$ in a symmetric way as
\beq
\alpha = \int_{\hat \Lambda}^{\infty} P(\Lambda | \mathbb{H}_0) \rm{d} \Lambda , 
\eeq
with $P(\Lambda \le \hat \Lambda | \mathbb{H}_1) = P(\Lambda \ge \hat \Lambda | \mathbb{H}_0)$. We convert $\alpha$ to a separation power $Z$ between two signal hypotheses with $Z=2\tilde Z$, where $\tilde Z$ is a significance (in $\sigma$)\,\cite{Chen:2013waa}.
We see that Riemannian preprocessing achieves the better separating power  even with same statistics as in Tab.\,\ref{tab:results}.
%%%%%%%%%%%%%%%%%
\begin{table}[h]
\begin{tabular}{|c|c|c|c|}
\hline
Preprocessing              & Phase space                       & $N_{\textrm{evt}}$ & Separating power ($\alpha$)\\ \hline
\multirow{2}{*}{Riemannian} & \multirow{2}{*}{Resolved}   &  $18$ & 3.11\,$\sigma$ (0.05968) \\  \cline{3-4}
                                              &                                             &  $219$ & 6.03\,$\sigma$ (0.00128) \\ \hline 
\multirow{2}{*}{Lund}            & \multirow{2}{*}{Merged}       &  $18$ & 2.55\,$\sigma$  (0.10134) \\ \cline{3-4}
                                              &                                              &  $219$ & 3.97\,$\sigma$ (0.02364) \\ \hline
\end{tabular}
\caption{Statistical significance in separating different hypotheses.
 }
\label{tab:results}
\end{table}
%%%%%%%%%%%%%%%%%

\noindent{\bf Conclusion} Even though automated feature engineering is the heart of ML,  we point out the importance of a data preprocessing in applying ML to the LHC analysis. 
As we observe, a smart data preprocessing method helps to focus on a unique feature more efficiently and to achieve a statistical stability
 in making a decision with a finite size of data. One can utilize our method in reducing QCD backgrounds to enhance a significance for the higgs related processes. We briefly mention about the improvement in reducing backgrounds in $HZ\to b\bar b \ell\bar \ell$ channel in appendix\,\ref{sec:appendix2}. It would be also interesting to compare our method with 
various decorrelation methods\,\cite{Chang:2017kvc,Shimmin:2017mfk, Bradshaw:2019ipy, Kasieczka:2020yyl}.

\section*{acknowledgments}
MP appreciates Chul Kim for introducing a Lund analysis.
This work is supported by NRF-2021R1A2C4002551. 
%%%%%%%%%%%%%%%%%%%
\newline
\appendix
\section{The structure of CNN} \label{sec:appendix1}

For each event we construct an image as  a square array in the ($\eta\text{-}\phi$) plane with each pixel intensity given by the total hadrons $p_T$ deposited in the associated region in the calorimeter. The rectangular region between $-2.5\le\eta\le2.5$ and $-\pi\le\phi\le\pi$ is discretized into $50\times 50$ pixels grid. As discussed earlier, in order to optimize the CNN performance and reduce the error, we applied the following preprocessign steps:
\begin{enumerate}
\item {\bf{Image cleansing:}} removing all leptons and photons from the image.
\item {\bf{Center:}} shift the center of the image from $(0,0)$ to $(\frac{\eta_{b^-}+\eta_{b^+}}{2},\frac{\phi_{b^-}+\phi_{b^+}}{2})$.  
\item {\bf{Normalization:}} normalize pixels intensity by diving each pixel in the image by the maximum pixel intensity value. 
\item {\bf{Inverse stereographic projection:}} project the image pixels in $(\eta\text{-}\phi)$ plane to a Riemann sphere by applying the inverse stereographic transformations. We fix the hot cores position in $\phi_R$ dimension to be at ($-\frac{\pi}{2},\frac{\pi}{2}$). The projected  images represent sphere in three dimensions and thus we cannot use the normal CNN to classify them. One way is to use a spherical CNN approach as presented in \cite{Perraudin:2018rbt}. Another way is to use the Mollweide projection as shown in FIG.\,\ref{fig:3} (left). In this work we use Mollweide projected images as input to the CNN.
\item {\bf{Momentum smearing:}} smear the momentum according to Gaussian distribution and correlate the neighbouring pixels to decrease the sparsity in images \cite{Buss:2022lxw}. Fig.\,\ref{fig:3} illustrates the effect of Gaussian filter with standard deviation $\sigma=3$ on the reconstructed images for  singlet, octet  scalars and backgrounds.
 
\end{enumerate}
\begin{figure}[t!]
\includegraphics[width=0.45\textwidth]{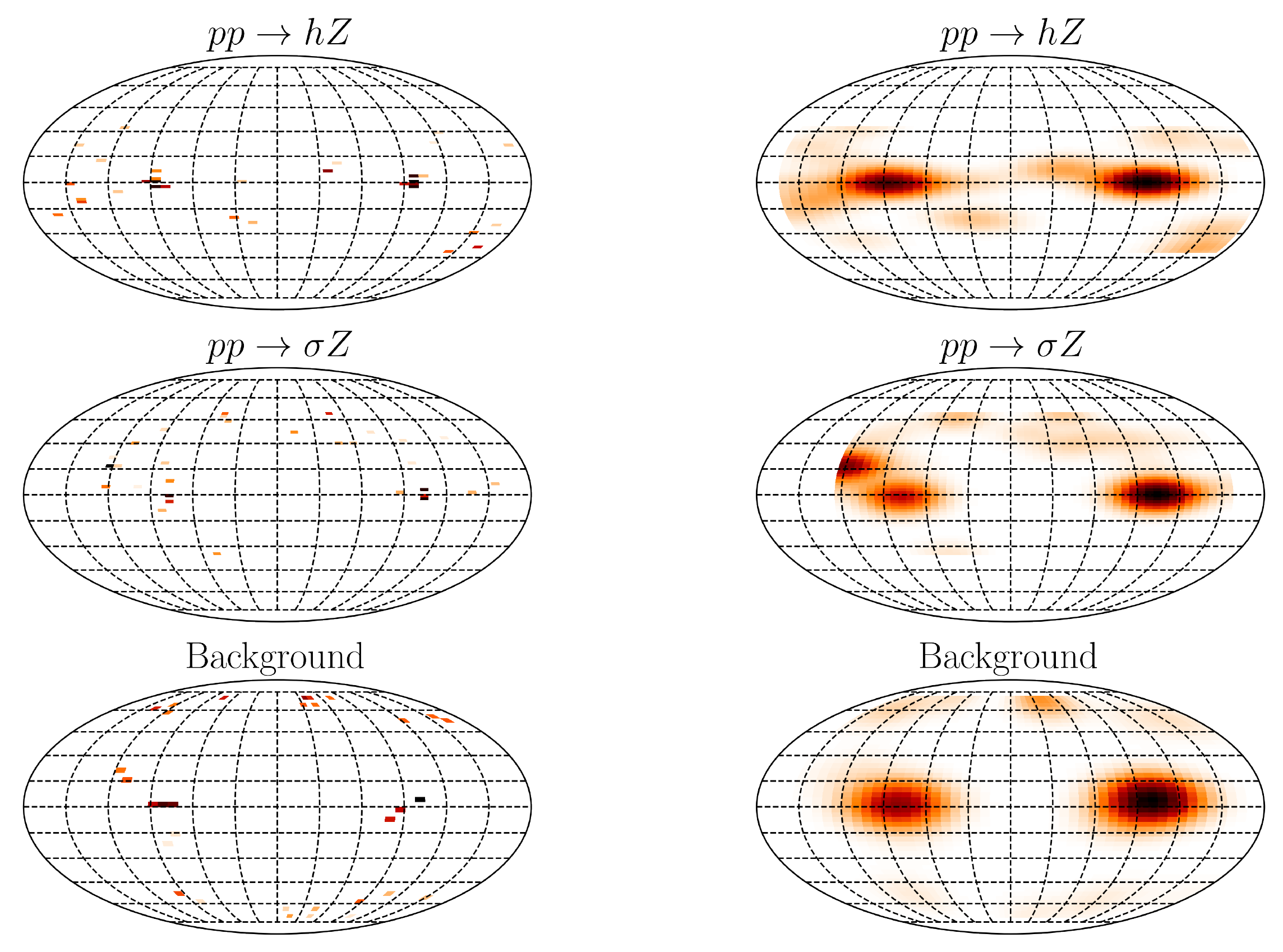}
\caption{Mollweide projection for a single event before (left) and after (right) applying Gaussian kernel with standard deviation $\sigma=3$.}
\label{fig:3}
\end{figure}

Testing the CNN classification performance on two different data sets, Riemannian preprocessing and No preprocessing, requires to  optimize the CNN structure for each data set  individually. We found that CNN with  six  convolution layers with kernel size 3, two dense layers and one output layer with two neurons is the best choice for both cases. Each  convolution layers pair is followed by a maxpooling layer of size 2 and a dropout layer with dropout rate of $20\%$. Dense layers are followed by dropout layer with dropout rate of $30\%$. The number of kernels in the first convolution layer is fixed to 16 and  the activation functions is ReLU except the last dense layer (output layer)  we use SoftMax function. To maintain the network stability and to avoid the covraiate  shift problem we add batch normalization layers. In Riemannian preprocessing images,  edges are of most important thus we apply padding layer in the first convolution layer to keep image dimension intact. The Loss function, categorical cross entropy, is minimized using Adam optimizer with learning rate of $0.001$. The number of kernels in the convolution layers and the number of neurons in the dense layers has been optimized using RandomizedSearchCV function in Sikit-learn. We use a balanced data set of 100K events for both processes, each data set is divided into $60\%$ training set, $20\%$ validation set and $20\%$ test set.

\section{Background rejection} \label{sec:appendix2}

\begin{figure}[h!]
\includegraphics[width=0.4\textwidth]{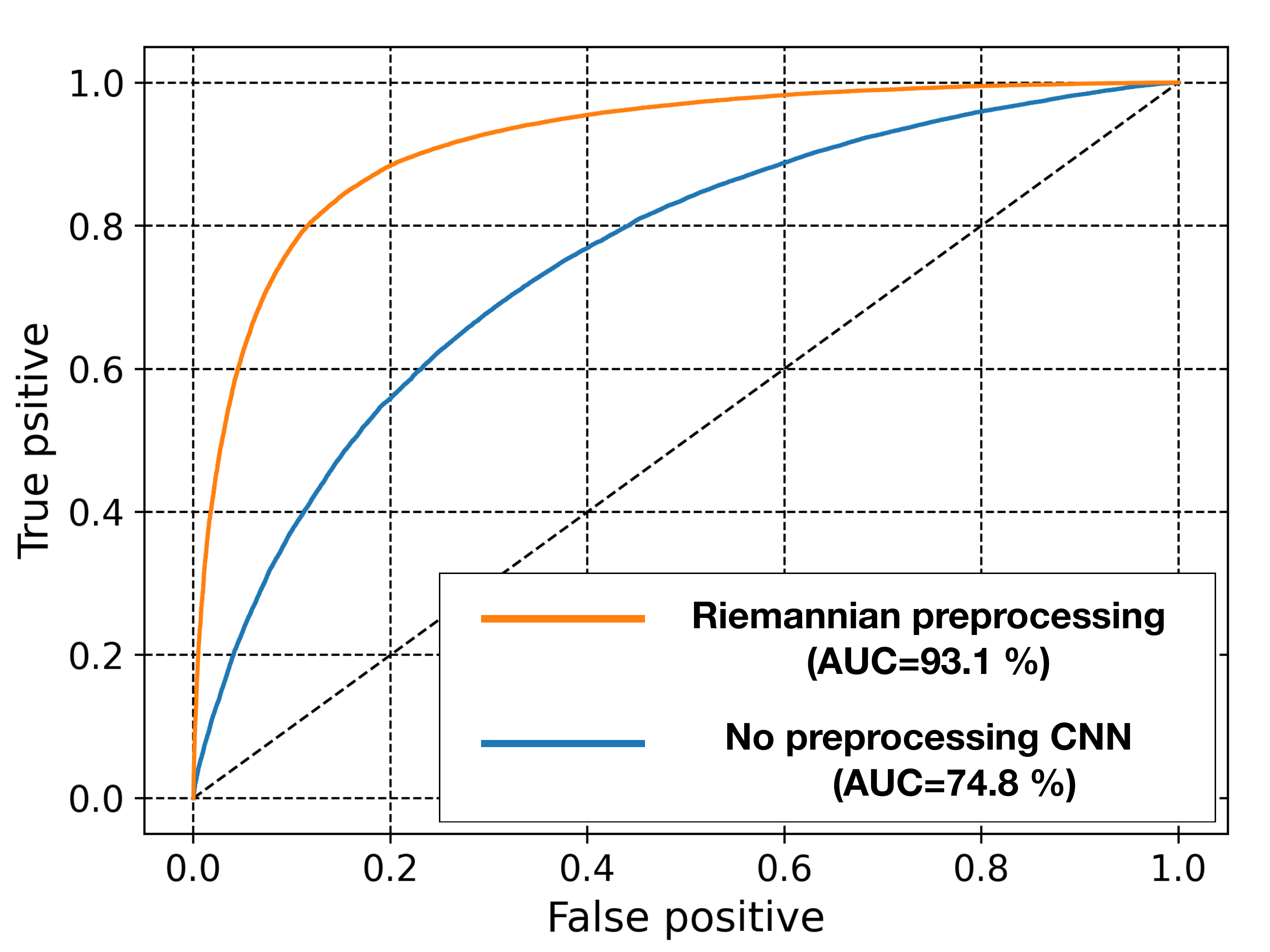}
\caption{ROC curve in separating backgrounds from signal events by utilizing color flow information in a resolved phase space. }
\label{fig:ROCBKG}
\end{figure}

Background contribution in a $HZ\to b \bar b \ell \bar \ell$ channel comes from vector boson plus three or more jets, di-top production, single top production and di-boson production\,\cite{ATLAS:2020fcp}; while we dropped the $W+jets$ and $Wt$ processes as they have negligible contribution after passing all selection cuts. 
In \cite{ATLAS:2020fcp}, ATLAS analysis does not take an advantage of different color flow information between a signal and backgrounds. Here we add a CNN study as in the main text with / without Riemannian preprocessing. In FIG.\,\ref{fig:ROCBKG} we present ROC performance using CNN analyses. We can achieve an extra factor 2 enhancement in a discovery significance using Riemannian preprocessing while we get only $25\%$ enhancement with a normal CNN analysis when we take a false positive as 0.2 for example.

%%%%%%%%%%%%%%%%%%%
\bibliography{reference}
%%%%%%%%%%%%%%%%%%%
\end{document}